\documentclass[10pt,letterpaper,pra,twocolumn,superscriptaddress]{revtex4-1}
\usepackage{graphicx,color}
\usepackage{bm}
\usepackage{float}
\usepackage{ulem}
\usepackage{amsmath}
\usepackage[utf8]{inputenc}
\usepackage[colorlinks=true,linkcolor=blue,citecolor=magenta,urlcolor=blue]{hyperref}

\newcommand{\ket}[1]{\ensuremath{|#1\rangle}}
\newcommand{\bra}[1]{\ensuremath{\langle#1|}}
\makeatletter
\renewcommand*\env@matrix[1][c]{\hskip -\arraycolsep
  \let\@ifnextchar\new@ifnextchar
  \array{*\c@MaxMatrixCols #1}}
\makeatother
\begin{document}

\title{Multi-dimensional entanglement generation with multi-core optical fibers}

\author{E.~S.~Gómez}
\email{estesepulveda@udec.cl}
\affiliation{Departamento de F\'isica, Universidad de Concepci\'on, 160-C Concepci\'on, Chile}
\affiliation{Millennium Institute for Research in Optics, Universidad de Concepci\'on, 160-C Concepci\'on, Chile}

\author{S.~Gómez}
\author{I.~Machuca}
\affiliation{Departamento de F\'isica, Universidad de Concepci\'on, 160-C Concepci\'on, Chile}
\affiliation{Millennium Institute for Research in Optics, Universidad de Concepci\'on, 160-C Concepci\'on, Chile}

\author{A.~Cabello}
\affiliation{Departamento de F\'{\i}sica Aplicada II, Universidad de Sevilla, E-41012 Sevilla, Spain}
\affiliation{Instituto Carlos~I de F\'{\i}sica Te\'orica y Computacional, Universidad de Sevilla, E-41012 Sevilla, Spain}

\author{S.~Pádua}
\affiliation{Departamento de Física, Universidade Federal de Minas Gerais, 31270-901 Belo Horizonte, Minas Gerais, Brazil}

\author{S.~P.~Walborn}
\author{G.~Lima}
\affiliation{Departamento de F\'isica, Universidad de Concepci\'on, 160-C Concepci\'on, Chile}
\affiliation{Millennium Institute for Research in Optics, Universidad de Concepci\'on, 160-C Concepci\'on, Chile}

\begin{abstract}
Trends in photonic quantum information follow closely the technical progress in classical optics and telecommunications.  In this regard, advances in multiplexing optical communications channels have also been pursued for the generation of multi-dimensional quantum states (qudits), since their use is advantageous for several quantum information tasks. One current path leading in this direction is through the use of space-division multiplexing multi-core optical fibers, which provides a new platform for efficiently controlling path-encoded qudit states. Here we report on a parametric down-conversion source of entangled qudits that is fully based on (and therefore compatible with) state-of-the-art multi-core fiber technology. The source design uses modern multi-core fiber beam splitters to prepare the pump laser beam as well as measure the generated entangled state, achieving high spectral brightness while providing a stable architecture. In addition, it can be readily used with any core geometry, which is crucial since widespread standards for multi-core fibers in telecommunications have yet to be established. Our source represents an important step towards the compatibility of quantum communications with the next-generation optical networks.
\end{abstract}
\maketitle

\section{Introduction}
In quantum information, there are several protocols that have an improved performance when implemented with multi-dimensional quantum systems (qudits). For instance, single qudit states can be exploited for building quantum cryptographic schemes supporting more component imperfections \cite{Cerf}, for efficient strategies solving communication complexity problems \cite{Armin_2017,Armin_2018}, and for advanced phase-estimation algorithms  \cite{Araujo_2014,Taddei_2020}. Moreover, multi-dimensional entanglement allows for two critical advantages related with Bell tests of quantum non-locality \cite{bell}, which are the building blocks of entanglement-based quantum information protocols \cite{bellrmp}. Specifically, some Bell inequalities for qudits have the property that their genuine quantum violation can still be achieved while working with lower (compared to qubits) overall detection efficiencies \cite{Brunner_2010}, which is arguably the main technological challenge of loophole-free Bell experiments. Second, there is a family of Bell inequalities specially tailored for entangled qudits \cite{Zeilinger_2000,cglmp}, whose quantum violation can still be achieved in a regime where noisy setups would be regarded as useless if based solely on the famous Clauser-Horne-Shimony-Holt Bell inequality  \cite{CHSH,ClauserHorne}.

Traditionally, the transverse momentum of single photons has been used for encoding qudit states for over almost two decades now \cite{Padgett_2002,Zeilinger_2002,Neves_2005,Boyd_2005,Steve_2006}. The distribution of photonic quantum states using optical fibers is a
fundamental building block towards quantum networks, but due to effects such as decoherence-inducing mode coupling and limited amount of modes supported, the transmission of such spatially encoded qudits over conventional multi-mode and single-mode fibers has always been considered a formidable challenge. Interesting strategies in this direction have been presented recently \cite{Mehul_2019,Forbes_2019}, but they are still limited to the transmission in the range of a few meters for multi-mode fibers, and unable to fully exploit the advantages provided by qudits in the case of single-mode fibers (SMF). Alternatively, there is a new trend emerging for the fiber propagation of spatially encoded qudits that is based on new types of optical fibers developed for space-division multiplexing (SDM) in classical telecommunications \cite{GuixReview_2019}. Basically, SDM technology increases data transmission in classical networks by adopting fibers capable of simultaneously supporting several transverse optical modes, where the information is then multiplexed \cite{Richardson_2013}. Since mode coupling in these fibers is minimal, high-fidelity coherent transmission of entangled and single spatially encoded qudits has been already demonstrated up to a few kilometers \cite{Cozzolino_2018,Cao_2018,Lio_2020}.

In this work we introduce a new source of entangled path-encoded qudit states, which is fully based on SDM multi-core fiber (MCF) technology. Multi-core fibers have several cores within the same cladding, and each core mode can be used to define the logical basis in the path encoding strategy, as we explain below. Moreover, the relative phase between two different cores of a MCF has been shown to be orders of magnitude more stable than two single-mode fibers \cite{Lio_2020}. Consequently, these fibers have already been exploited for multi-dimensional quantum cryptography \cite{Canas_2017,Ding_2017}, quantum random number generation \cite{Optica_2020}, quantum computation \cite{Taddei_2020}, and Bell inequality violation \cite{Lee_2017,Lee_2019}.  Path encoding has the important appealing that $d$-dimensional arbitrary operations can be implemented with conventional linear optical elements \cite{Reck}, and has now become the standard encoding strategy in experiments with integrated photonic circuits \cite{Obrien_2015,Bristol_2018}. Therefore, our source of path entanglement can be used as a resource for the implementation of entanglement-based quantum information protocols in experiments based solely on new MCF technology \cite{Optica_2020}, or also to efficiently distribute multi-dimensional entanglement between integrated circuits, which are proven to be compatible with MCFs \cite{Ding_2017}. Our source compares favorably in terms of integrability and brightness with previous works for fiber-based generation of multi-dimensional path entanglement \cite{Zeilinger_2012,Lee_2017,Lee_2019}, and represents an important step towards the compatibility of quantum communications with the next-generation SDM optical networks.

\section{Quantum information with MCF technology}
\begin{figure}
\centering
\includegraphics[width=6cm]{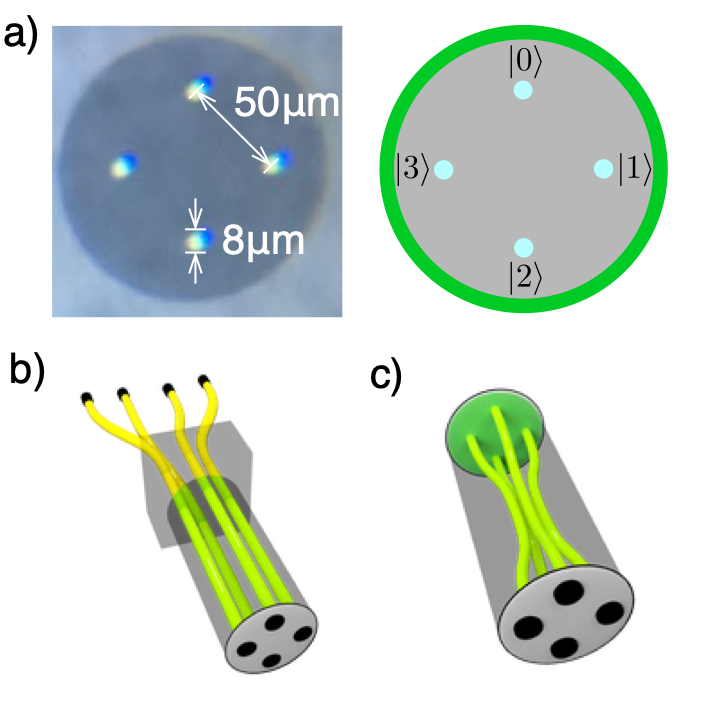}
\caption{a) Photo of a multi-core fiber with four cores (4CF) taken with a camera and a fiber microscope (left), and diagram of the path-encoding strategy for defining the logical states in dimension $d=4$ (right).  b) The demultiplexer device, coupling single-mode fibers to a multi-core fiber (here 4CF). c) A $4\times 4$ beam splitter constructed within a 4CF.}
\label{fig:core_crystal}
\end{figure}
Multi-core fiber is currently being pursued for its capability to increase communications rates in telecommunications \cite{Richardson_2013}.  At the same time, several authors have investigated MCF technologies in photonic quantum information platforms (see Ref. \cite{GuixReview_2019} for a review). Here we employ four-core fibers (4CFs).  Fig. \ref{fig:core_crystal} a) shows an image of the facet of a 4CF taken with a standard fiber-inspection microscope and a camera.  The cores are $\sim$8$ \mu$m wide, corresponding to single-mode at $1550$ nm, arranged on the vertices of a square with $50\mu$m sides, which is large enough so that cross-core coupling is greatly reduced.  This allows us to treat the spatial modes of each core as independent.  For the 4CF, we thus define the path logical basis consisting of the core states $\ket{j},(j=0,\dots,3)$, as also shown in Fig. \ref{fig:core_crystal} a).

In addition to patchcords of MCF, crucial MCF-compatible optical devices have been recently developed that will allow widespread use of MCF technology in photonic quantum information.  First is the demultiplexer (DM) device, as exemplified in Fig. \ref{fig:core_crystal} b), which allows one to couple $N$ single-mode fibers to an $N$-core fiber. With this, light can be sent from a standard single-mode fiber into one core of a MCF or vice versa, providing compatibility with standard optical fiber components.  For example, we can couple a photon in mode $\ket{j}$ of an MCF via a DM to a fiber-ready single-photon detector.

Another important device is the MCF-based beam splitter (MCF-BS), shown in Fig. \ref{fig:core_crystal} c) for a 4CF.  The beam splitter is produced by heating and stretching a section of homogeneous MCF (without refractive index trenches), so that the the cores become closer together, enabling evanescent coupling between the cores \cite{Ming_2016}. The 4CF-BS and a 7CF-BS were characterized experimentally in Ref. \cite{Optica_2020}. When the proximity region is small, one can achieve an approximate 25\% coupling between all four cores of a 4CF \cite{Optica_2020}. In the experiment reported here, we use two types of 4CF-BS devices. The first has been designed for use at 775 nm, and the second at 1550 nm. More technical details are given below.  To reasonable approximation, the 4CF-BS can be represented by the unitary matrix
\begin{equation}U_{BS}=\frac{1}{2}
\begin{bmatrix}[r]
 1 & 1 & 1 & 1 \\
 1 & 1 & -1 & -1 \\
 1 & -1 & 1 & -1 \\
 1 & -1 &  -1 & 1
  \label{Matrix_Theo}
\end{bmatrix}.\end{equation}
The 4CF-BS thus takes a photon in path state $\ket{j}$ to an equally-weighted superposition state of the form $\ket{\psi_j}=\frac{1}{2} \sum_{k}u_{kj}\ket{k}$, where $u_{kj}=\pm 1$ are the entries of the matrix \eqref{Matrix_Theo}. Likewise, the 4CF-BS can be used to map superposition states into logical basis states: $U_{BS} \ket{\psi_j}=\ket{j}$.

These MCF devices can be connected with relatively low losses (about 2-5\% at 1550 nm) using standard FC/PC fiber connectors \cite{Optica_2020}. We note that the future development of one more MCF-based optical primitive, namely MCF-integrated phase shifters, will allow for entire multi-path optical circuits to be created entirely within multi-core fibers.  This will allow for implementation of complex interferometers in a relatively stable MCF platform \cite{Lio_2020}, for use in both quantum and classical optics applications such as communications and metrology.  Moreover, multi-port beam splitter devices have recently been shown to provide advantages in quantum logic operations \cite{Saygin_2020,PUMA_2020}. This motivates the development of entangled-photon sources that are compatible with MCF technology.

\section{Experiment}
\subsection{Setup}\label{sec:MS}
\begin{figure*}[t]
\centering
\includegraphics[width=0.75\textwidth]{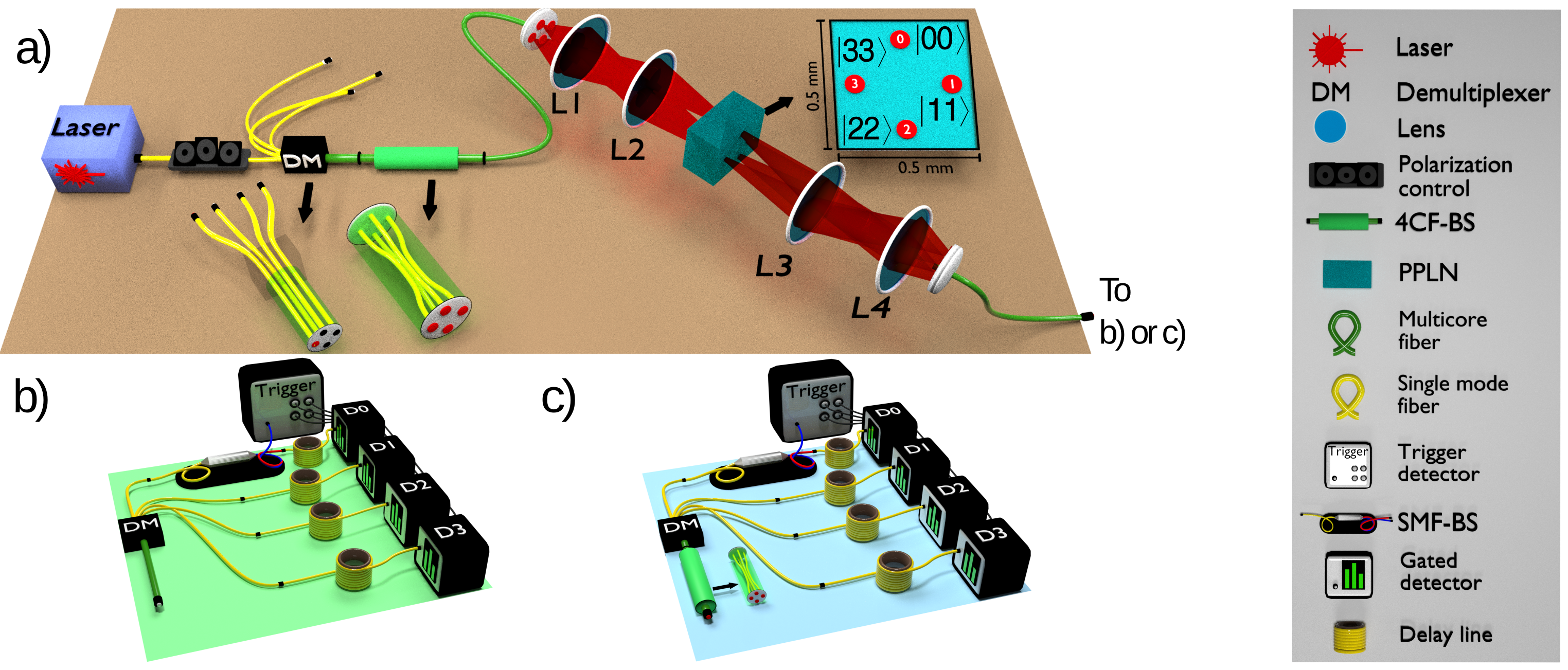}
\caption{a) Schematics of the experimental setup. Entangled qudit states are generated by resorting to coherent illumination of multiple regions of a non-linear crystal and MCF technology. The generated entangled qudits are sent to one of the detection systems. b) Detection system for $Z$ measurements, in which each core of the source output MCF is coupled to a SMF via a demultiplexer DM.  Single photon detectors register coincidences between the photon pairs. c) The detection system for $X_j$ basis measurements. In this case, the output of the entanglement source is first coupled to a $4\times4$ MCF-BS and its output fiber is the coupled to SMFs using a DM.}
\label{setupHD}
\end{figure*}

Spontaneous parametric downconversion (SPDC) is a non-linear optical process that has been traditionally used to produce correlated photon pairs \cite{klyshko69,burnham70}. Here we use SPDC to construct a source of entangled path qudit states, which is fully based on (and therefore compatible with) state-of-the-art MCF technology. The adopted setup is show in Fig. \ref{setupHD} a). A fiber-based continuous-wave diode laser operating at $773$ nm is used to excite a 1cm long type-0 periodically poled lithium niobate (PPLN) crystal generating down-converted photon pairs at a center wavelength of $1546$ nm.  The primary goals for the design were for the source to be scalable, to have high spectral brightness and a stable performance.  The main idea is to coherently illuminate the PPLN crystal in four regions corresponding to the cores of the fiber, following the same core layout as that of the face of the 4CF shown in Fig. \ref{fig:core_crystal} a). We note that multi-spot coherent illumination of non-linear crystals for the generation of path-entangled photons was first presented in the 1990's \cite{Shi_1997} and has been used frequently since then \cite{Zeilinger_2012,Bristol_2018,Lee_2019,Guo_2020}. However, to our knowledge, our source is the first one where the preparation and measurements stages are fully built of new SDM optical fiber technology aiming for compactness and connectivity with next-generation optical fiber networks. To coherently illuminate our PPLN crystal in a robust and efficient way, we take advantage of the developed MCF beam splitters described above \cite{Optica_2020}. The pump laser was connected to a SMF of the DM, and consequently coupled to a single core of the 4CF at its output side. Then, the pump beam is sent through a custom made 4CF-BS, designed to operate at 775 nm, that coherently splits the pump beam among the four cores of the 4CF. The recorded split ratios of this 4CF-BS were $23.79\%$, $24.88\%$, $27.19\%$, and $24.14\%$ at 773 nm. A fiber polarization controller (PC), placed before the DM, was used to guarantee that the pump polarization could be aligned with the extraordinary axis of the PPLN crystal, to maximize the photon pair generation rate.

The output face of the 4CF-BS was imaged onto a plane at the center of the PPLN crystal using lenses $L_1$ (focal length $f=11$ mm) and $L_2$ ($f=50$ mm), arranged confocally, giving a $\sim$$4.5\times$ magnified image of the 4CF face. The down-converted pairs are produced at each one of the illuminated regions corresponding to the four fiber cores of the MCF (See the crystal inset of Fig. \ref{setupHD} a)). The generated photon pairs are then sent through confocal lenses $L_3$ ($f=50$ mm) and $L_4$ ($f=11$ mm) that perform the inverse operation: creating a de-magnified image of the down-converted pairs at the face of the output 4CF, coupling them into the fiber.  A bandpass and interference filter centered at $1540$ nm are used (between the lenses, not shown for sake of clarity) to remove the remaining light from the pump beam.

Considering the split ratio recorded of the 4CF-BS, and that it is not possible in principle to distinguish which region of the crystal produced the photon pair, the generated two-photon state can be approximately written as a coherent superposition of the form
\begin{equation}\label{eqn:Psi}
\vert \Psi \rangle =\dfrac{1}{2}(\vert 00 \rangle + \vert 11 \rangle + \vert 22 \rangle + \vert 33 \rangle),
\end{equation}
where we use the shorthand notation $\vert j j  \rangle=\ket{j}_A\otimes \ket{j}_B$.

After the source, the down-converted photons are sent to one of two measurement devices shown in Fig. \ref{setupHD} b) and Fig. \ref{setupHD} c). In each of these, photon pairs were detected by measuring coincidence counts $C_{jk}\,(j,k=0,1,2,3)$, where here $j$ and $k$ refer to the core modes of the down-converted idler and signal photon, respectively.   To detect the photons in the path ($Z$) basis, we used the measurement system sketched in Fig. \ref{setupHD} b).  The output 4CF of the source is connected to a DM, coupling each core to a SMF. To detect photons propagating over the SMFs a home-made coincidence count system is used, which works as follows: a free-running trigger single photon detector (Idquantique ID220) is connected to one of the SMFs, let's say the fiber ``0'' (associated to core ``0'') as shown in Fig. \ref{setupHD} b), using a standard beam splitter (SMF-BS). When it registers a photon, a sync electrical pulse is sent to the four externally gated detectors (Idquantique ID210) for coincidence detections. These detectors are connected to the SMFs using optical delay lines such that coincidence counts can be registered using the sync signal. The ID210 detectors were configured with $25\%$ detection efficiency, and $5$ ns gate width, while the ID220 detector was configured with $15\%$ detection efficiency and $5\,\mu s$ of dead time. To observe all of the sixteen possible coincidence events $C_{jk}$, the ID220 was connected through a SMF-BS to each of the four SMFs, and counts of the form $C_{j0}, C_{j1}, C_{j2}, C_{j3}$ were recorded for all values of $j$ with an integration time of 5 s.

For measurements in bases complementary to the $Z$ basis, we first connected the output 4CF of the source to a 4CF-BS, as shown in Fig. \ref{setupHD} c). The MCF output of the 4CF-BS was then routed through a DM and each core coupled again to the SMFs.  The coincidence detection scheme was the same as in Fig. \ref{setupHD} b).  Including the relative phases corresponding to propagation in each core, the 4CF-BS allows us to measure in superposition bases of the form
\begin{subequations}
\label{eq:bsbasis}
\begin{align}
|0\rangle_D=&\frac{1}{2}\left(e^{i \phi_0} |0\rangle + e^{i \phi_1} |1\rangle+e^{i \phi_2} |2\rangle + e^{i \phi_3} |3\rangle \right),\\
|1\rangle_D=&\frac{1}{2}\left(e^{i \phi_0} |0\rangle + e^{i \phi_1} |1\rangle-e^{i \phi_2} |2\rangle - e^{i \phi_3} |3\rangle \right),\\
|2\rangle_D=&\frac{1}{2}\left(e^{i \phi_0} |0\rangle - e^{i \phi_1} |1\rangle+e^{i \phi_2} |2\rangle - e^{i \phi_3} |3\rangle \right),\\
|3\rangle_D=&\frac{1}{2}\left(e^{i \phi_0} |0\rangle - e^{i \phi_1} |1\rangle-e^{i \phi_2} |2\rangle + e^{i \phi_3} |3\rangle \right),
\end{align}
 \end{subequations} where $|j\rangle_D$ refers to the state at detector $Dj$ after the 4CF-BS, and $\phi_j$ are the relative phases acquired over core $j$.  One can switch between four mutually unbiased bases $X_j$, by choosing different values of the phases $\phi_j$.    We define the bases as $X_0 \rightarrow \phi_0=\phi_1=\phi_2=\phi_3=0$, $X_1 \rightarrow \phi_0=0$, $\phi_1=\pi$, $\phi_2=\phi_3=\frac{\pi}{2}$,  $X_2 \rightarrow \phi_0=0$, $\phi_1=\phi_3=\frac{\pi}{2}$, $\phi_2=\pi$, $X_3 \rightarrow \phi_0=0$, $\phi_1=\phi_2=\frac{\pi}{2}$, $\phi_3=\pi$. The $X_j$ bases are mutually unbiased to the $Z$ basis and also to each other.

\subsection{Source Characterization}
\begin{figure}[t]
\centering
\includegraphics[width=0.48\textwidth]{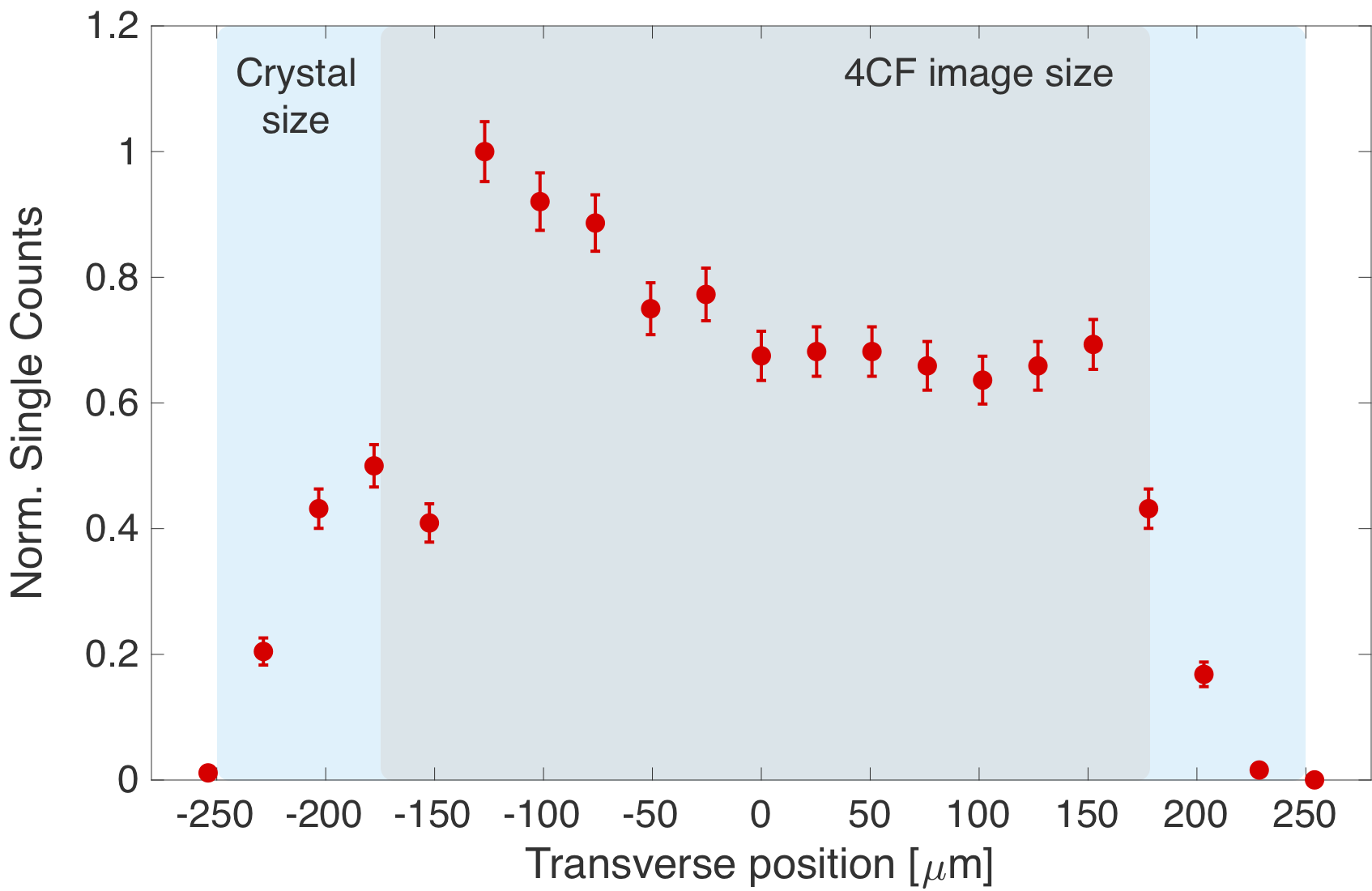}
\caption{Normalized single-counts while scanning the vertical ($y$-axis) transverse direction of the PPLN crystal recorded by a point-like single-photon detector.  The shaded regions correspond to the dimensions of the crystal and the image size of the 4CF.  \label{cuentasXY}}
\end{figure}

Now we describe several characteristics of the source. To ensure the spatially {homogeneous} generation of down-converted photons over the entire {transverse profile} of the crystal, {we performed a separate experiment in which} the 773 nm laser {was sent through a} single-mode fiber {and} used to illuminate the center of the crystal through the imaging system composed by lenses $L_1$ and $L_2$. The {down-converted light was} coupled into a point-like detector, composed {of} another single-mode fiber for $1550$ nm connected directly to an ID210 detector. We scanned {the crystal} along the {(transverse)} horizontal {($x$)} and vertical {($y$)} axis, recording the corresponding detector single-counts. As an example, Fig. \ref{cuentasXY} shows the spatial distribution of the down-converted photons with the crystal scanned in the $y$-direction. Similar results were obtained for the $x$-direction. We recognize a $\sim$$380$ $\mu$m width region where the photons can be created. Considering the core size, and the separation between them, we chose the  magnification factor $4.5$ such that the image size of the 4CF at the PPLN crystal was $\sim$$350$ $\mu$m. In this way, {the crystal} generates photon pairs in the four different illuminated regions defined by the image of the 4CF.

\begin{figure}[t]
\centering
\includegraphics[width=0.48\textwidth]{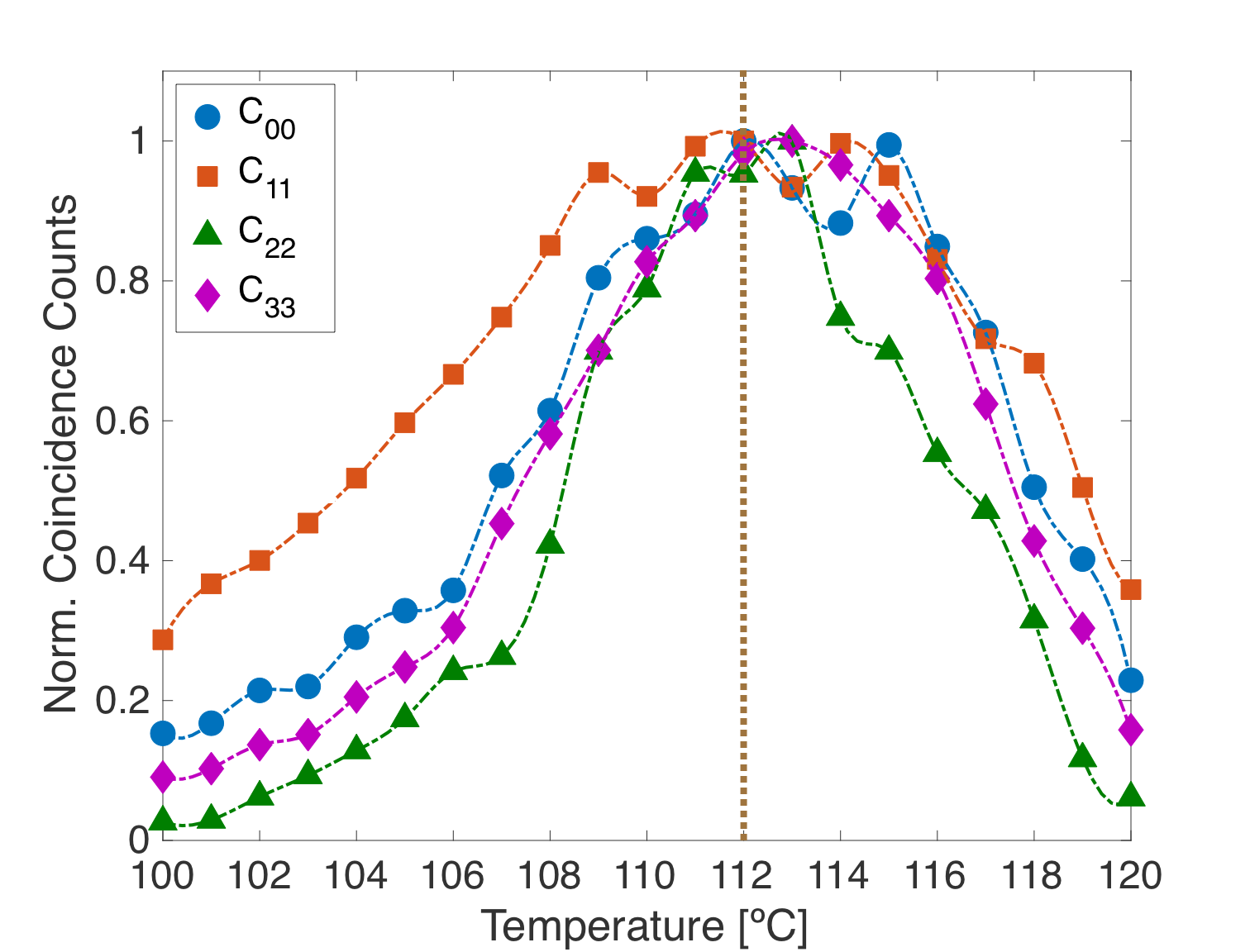}
\caption{Coincidence rate in each core as a function of the crystal temperature. The optimal temperature to satisfy the phase-matching conditions for all cores is 112$^\circ$C. \label{ccvsT}}
\end{figure}

Further, {since quasi-phase matching conditions are achieved by adjusting the temperature of the crystal, we tested that the optimal temperature was the same for each of the regions corresponding to the four cores}. We adjusted the PPLN crystal temperature {by placing it in} an electronically-controlled oven. Fig. \ref{ccvsT} shows normalized coincidence counts in each core, as a function of the temperature, recorded using the measurement scheme of Fig. \ref{setupHD} b). We observe that at 112$^\circ$C, a nearly optimized coincidence rate of SPDC in all cores was reached, in correspondence with the pump and down-converted photon wavelengths and the crystal poling period (19.8 $\mu$m).

Brightness of the source is an important characteristic, if entanglement is to be distributed over long-distances through optical fibers. Our entanglement source operates with a pump laser power of 1 mW per core. Taking this into account, plus the insertion losses of the optical devices and detector efficiencies ($\sim$4\% in coincidence), the observed spectral brightness {of the source} was $\sim 350000$ photon pairs (s mW nm)$^{-1}$, which is comparable to optimized sources for polarization entangled photons \cite{gomez16PRL}. Considering the typical loss in multi-core optical fibers of 0.4 dB per km for $1550$ nm, it would be possible to distribute multi-dimensional entanglement over at least $\sim$75 km {of fiber}.  Thus, this source can be readily employed to investigate the propagation of spatial entanglement over long multi-core fibers, which has yet to be realized.

Another remarkable feature of our source is the {phase stability provided by coupling the pump and down-converted light directly in and out of 4CFs, which are inherently robust against thermal fluctuations and mechanical stress since the cores lie within the same cladding \cite{Canas_2017,Lio_2020}. The relative phase between two different cores of a MCF can be at least two orders of magnitude more stable than two single-mode fibers over a 2km length \cite{Lio_2020}. When the $X_j$ measurement bases are implemented, the last 4CF-BS combines together the incoming down-converted photons, {forming an interferometer that is} sensitive to phase fluctuations. Thanks to the stability of the 4CF, without any fiber isolation system {or active phase stabilization}, we observe that phase {oscillations typically occur with a period of} several minutes. Fig. \ref{stab} shows the absolute value of the frequency spectrum for sample coincidence measurements, recorded at different times, in a \textsl{busy} laboratory environment. The peaks of these distributions correspond to oscillations with period between $2$ and $6.5$ minutes. We can see that there are essentially no frequencies components above $0.008$ Hz. Our goal here is to present the fundamental characteristics of the source but we note that, if necessary, one can employ the technique of Ref. \cite{Optica_2020} to phase-lock the source.

\begin{figure}[t]
\centering
\includegraphics[width=0.48\textwidth]{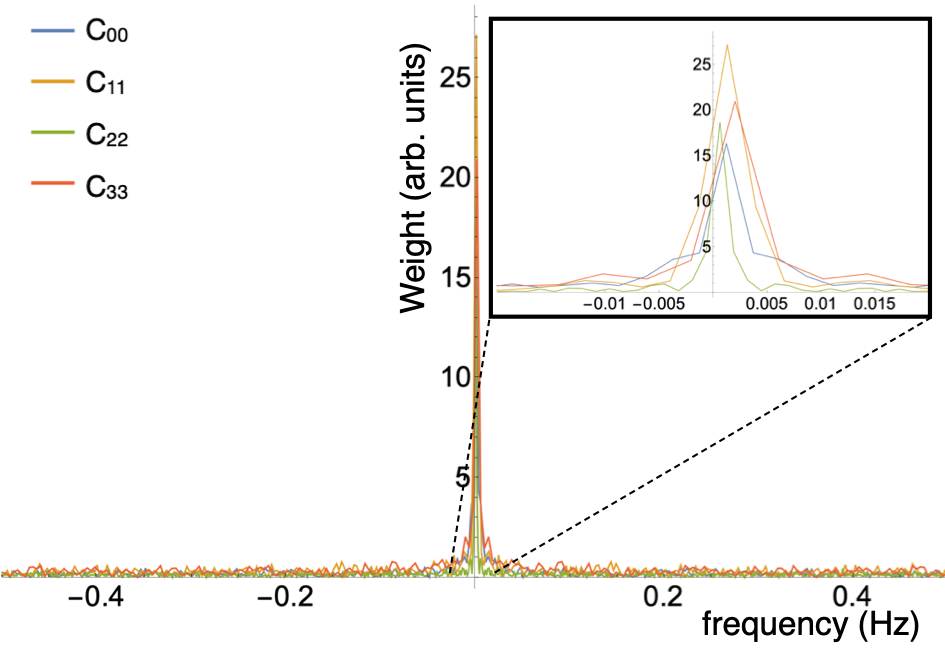}
\caption{Spectrum of frequency components of the coincidence counts between different detectors using detection system \ref{setupHD} c), taken at different times.  Fluctuations due to phase changes occur on a time scale of several minutes.}\label{stab}
\end{figure}

\subsection{Entanglement Certification}
We measured coincidence counts $C_{jk}$ as was explained in section \ref{sec:MS}. Accidental coincidence counts $a_{jk}$ arising from dark counts and ambient light were evaluated by recording the average count rate with a large relative electronic delay between detectors.  The corrected coincidence counts, given by $C^{c}_{jk}=C_{jk}-a_{jk}$, were used to estimate the joint probability distributions: $P_{jk}=C^c_{jk}/\sum_{jk}C^c_{jk}$. Experimental error was calculated by assuming Poissonian count statistics and Gaussian error propagation. The recorded probability distributions while measuring both photons in the $Z$ basis is shown in Fig. \ref{prob} a), with photons always found in the same core.

For measurements in the $X_j$ bases, there are two-photon coalescence effects on the 4CF-BS that impose a convenient symmetry to the coincidence counts. By  considering the state of Eq. \eqref{eqn:Psi} and the measurements of Eqs. \eqref{eq:bsbasis}, we see that the coincidence count probabilities at detectors $j$, $k$ ($j,k=0,1,2,3$) are given by
\begin{equation}
P_{jk}^{the}=\left | \frac{1}{8} \sum_{m=0}^3 u_{mj}u_{mk} e^{i2\phi_m} \right |^2,
\label{eq:Pmn}
\end{equation}
where $u_{mj},u_{mk}=\pm1$ are the sign of the entries in the matrix \eqref{Matrix_Theo}. In this case, when the relative phases are such that complete constructive/destructive interference occurs, only 4 of the 16 probabilities $P_{mn}$ are non-zero.  For example, when all of the phases are equal, the four probabilities $P_{00}=P_{11}=P_{22}=P_{33}=1/4$.  Since these four probabilities sum to one,  all the rest are zero.  This case corresponds to detection of both photons in the $X_0$ basis.  For the phases corresponding to the $X_1$ basis, only probabilities $P_{01}=P_{10}=P_{23}=P_{31}=1/4$ are non-zero.  Likewise, for the $X_2$ basis we have $P_{02}=P_{20}=P_{13}=P_{31}=1/4$ and all others equal to zero, while for $X_3$ we have $P_{03}=P_{30}=P_{12}=P_{21}=1/4$ with the rest equal to zero.  In this fashion, observing a maximum in one coincidence count group while observing all of the others near zero allows us to identify the relative phase values.

The measurements in the $X_i$ bases were identified using the coincidence count signatures described above, while applying a controlled bending to the 4CF fiber at the output of the source, causing mechanical stress that changes the relative phases between cores \cite{Sergi_2019}. The joint probability distributions obtained from the four $X_i$ measurements are shown in Fig. \ref{prob}. The similarity, ${C_B}$, of the recorded and theoretical probability distributions can be quantified using the Bhattacharyya coefficient \cite{Bhatta}. In our case, it reaches ${C_B}=0.91 \pm 0.02$, showing good agreement between theory and experiment.

\begin{figure}[t]
\centering
\includegraphics[width=0.48\textwidth]{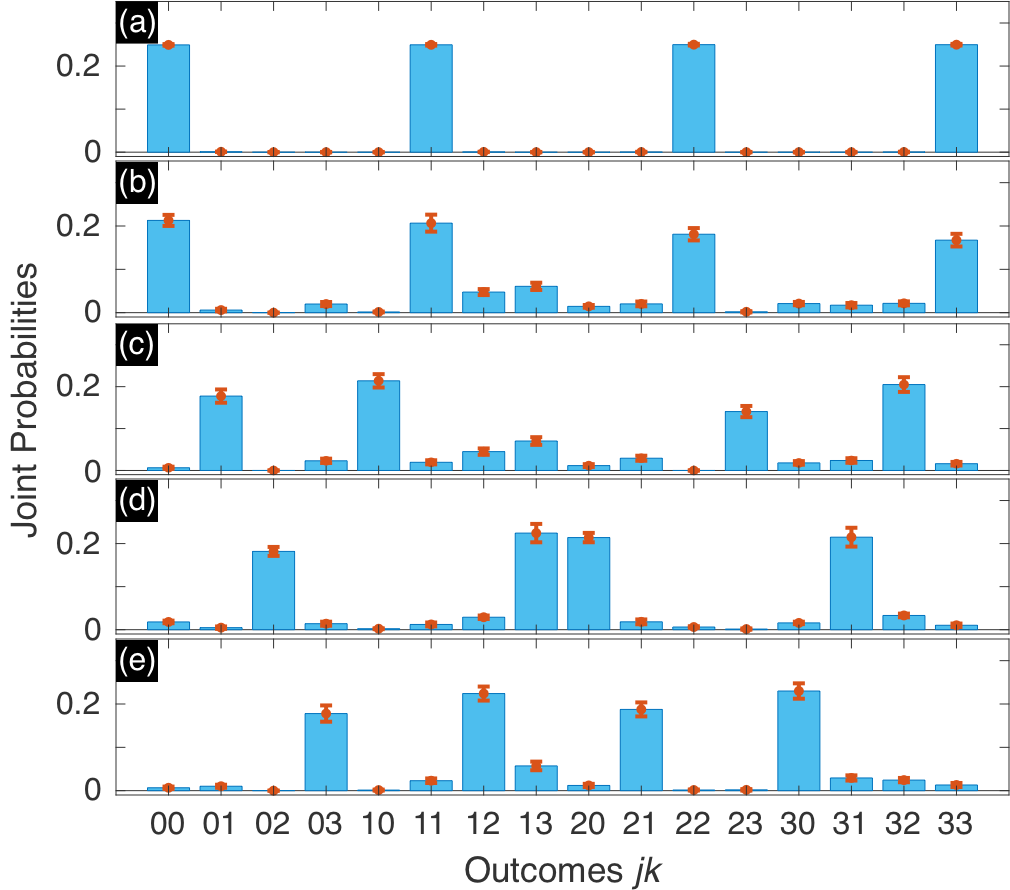}
\caption{Joint probability distributions obtained while measuring both photons in the (a) $Z$ basis, (b) $X_0$ basis, (c) $X_1$ basis, (d) $X_2$ basis, and (e) $X_3$ basis. Error bars are obtained considering Gaussian error propagation and Poissonian photo-count statistics. \label{prob}}
\end{figure}

\subsubsection{Fidelity}
To certify the multidimensional entanglement of the generated state $\rho$, we use its fidelity $F(\rho, \ket{\Psi})= \bra{\Psi} \rho \ket{\Psi} $ to the target state $\ket{\Psi}$ in Eq. \eqref{eqn:Psi}. Explicitly,
 \begin{equation}
F(\rho, \ket{\Psi})  = \sum_{j=0}^3 \langle jj | \rho | jj \rangle   +   2 \sum_{\substack{j=0 \\ k=j+1}}^3 \mathrm{Re} \left [ \langle jj | \rho | kk \rangle \right ].
\label{eq:Fid}
\end{equation}
The first term can be obtained directly from the measurements in the $Z$ basis, where we have $P_{jj} =  \langle jj | \rho | jj \rangle$. The second term corresponds to the coherence between states $\ket{jj}$ and $\ket{kk}$.  These can be calculated by defining Pauli operators for the $j,k$ subspace as \cite{Padua_2014,bavaresco18}: $\sigma^{(jk)}_x=\ket{j}\bra{k}+\ket{k}\bra{j}$, $\sigma^{(jk)}_y=i\ket{j}\bra{k}-i\ket{k}\bra{j}$. Then,  $\mathrm{Re} [ \langle jj | \rho | kk \rangle ] = (\langle \sigma^{(jk)}_x \otimes \sigma^{(jk)}_x\rangle - \langle \sigma^{(jk)}_y \otimes \sigma^{(jk)}_y \rangle)/4$.  These expectation values can be evaluated directly from the $X_i$ measurements, as we describe in detail in the Appendix. Using the recorded data of the 5 mutually unbiased measurements, we obtain $F=0.789 \pm 0.007$.  Since any state with $F(\rho, \ket{\Psi}) > 3/4$ is incompatible with a Schmidt number $ \leq 3$  \cite{Padua_2014,bavaresco18}, we can confirm the four-dimensional nature of the entanglement produced by the source.

\subsubsection{High-Dimensional Steering}

\begin{figure}[t]
\centering
\includegraphics[width=0.47\textwidth]{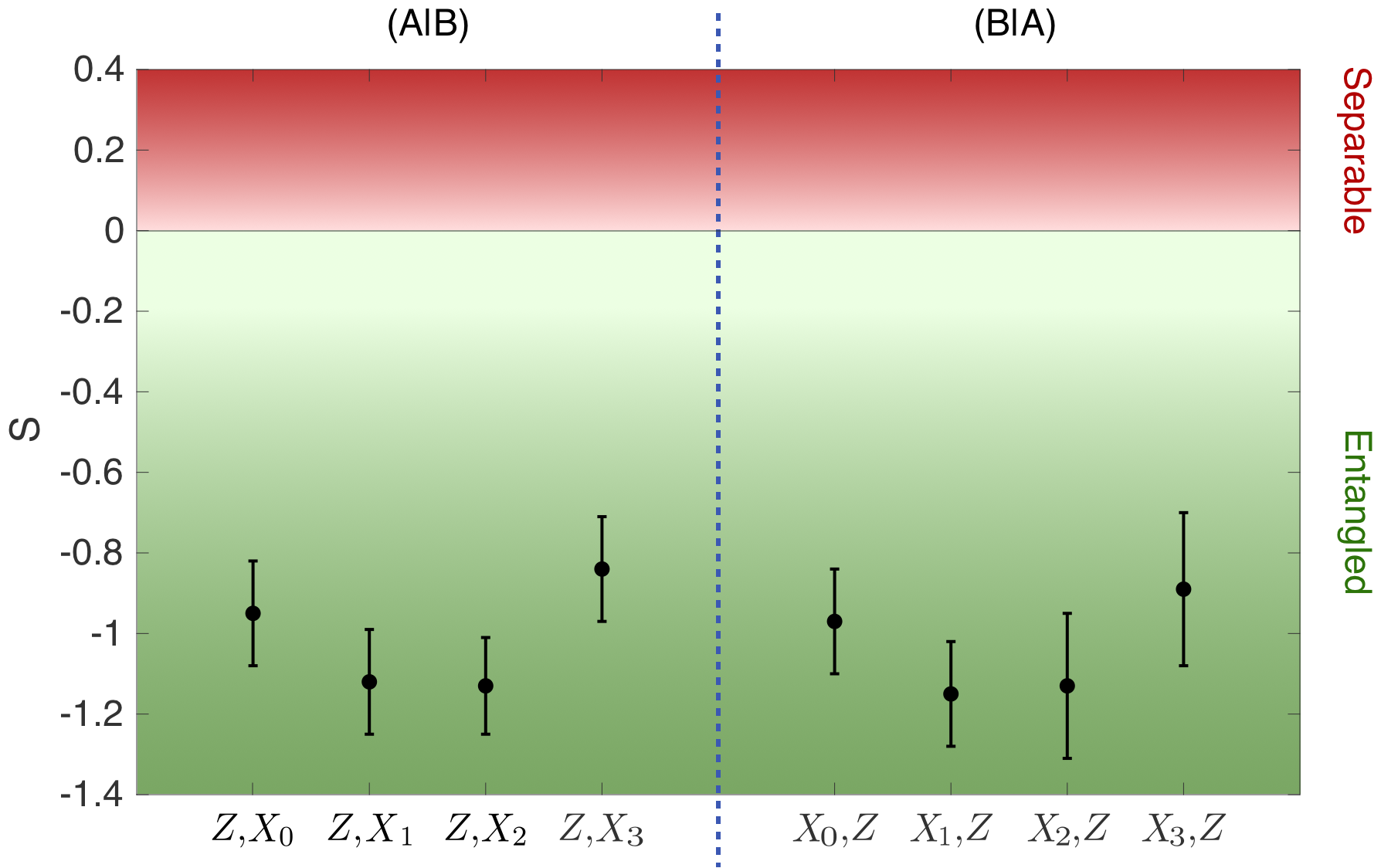}
\caption{Experimental values for the EPR-steering criterion $S$ for steering from $A$ to $B$, denoted $(B|A)$ and from $B$ to $A$, denoted $(A|B)$.  When $S_{JK}^{(PQ)}<0$ we can confirm that the state is entangled in the one-sided device independent scenario. \label{entropic}}
\end{figure}

Another interesting approach to certify multi-level entanglement generation is the one based on quantum steering \cite{wiseman07}, which is a distinct correlation that lies between entanglement and Bell-nonlocality, and has been related to one-sided device independent quantum cryptography \cite{branciard12}, as well as one-sided device independent randomness generation \cite{law14,passaro15,mattar17,Gomez_2018}. For two mutually unbiased bases corresponding to observables $\hat{P}$ and $\hat{Q}$, quantum steering can be identified using the entropic criterion \cite{schneeloch13}
\begin{equation}
S_{JK}^{(PQ)} = H(P_J|P_K) + H(Q_{J}|Q_{K}) - \log_2 D \geq 0,
\label{eq:Hcriteria}
\end{equation}
where $J,K=A,B$ ($J\neq K$) denote the subsystems.  $H(P_J|Q_K)$ is the conditional Shannon entropy calculated over the joint probabilities associated to measurements in the $P$ and $Q$ bases on subsystems $J$ and $K$, respectively. Steering can be an asymmetrical correlation, and violation of inequality \eqref{eq:Hcriteria} indicates steering from subsystem $K$ to subsystem $J$  when $S_{JK}^{(PQ)}<0$, which also indicates that the two systems are entangled. Identifying $P=Z$ and $Q=X_j$, we applied these inequalities to our experimentally obtained probabilities of Fig. \ref{prob}. The conditional entropy was calculated using $H(P_J|Q_K) = H(P_J,Q_K)-H(Q_K)$, where $H(P_J,Q_K)$ is the joint Shannon entropy and $H(Q_K)$ is the marginal entropy corresponding to  local measurement $Q$ on party $K$. The results are shown in Figure \ref{entropic}. We obtain negative values for all correlations tested, showing that the generated qudit state exhibits steering from $A$ to $B$ as well as from $B$ to $A$. We obtained mean values $\bar{S}_{AB}=-1.01\pm 0.06$ and $\bar{S}_{BA}=-1.04\pm0.08$. Identification of steering allows us to confirm that the generated state is entangled in a one-sided device independent scenario \cite{cavalcanti15}, meaning that full knowledge of the inner workings of one of the system's devices is not necessary. Moreover, the dimensionality of each subsystem can be observed through the Shannon entropies  $H(P_J)$ and $H(Q_J)$ of the marginal distributions, which are all very close to the maximum value of $\log_2 D = 2$ bits for $D=4$ dimensional subsystem. The mean value is $1.99 \pm 0.02$ bits, which is twice the limit of qubit subsystems.

\section{Conclusion}

In this work we present a new design of a source of multi-dimensional photonic entanglement that is based on space-division multiplexing optical fibers, thus being fully compatible with next-generation telecommunications fiber networks. The source uses state-of-the-art multi-core fiber technology to illuminate a non-linear crystal, and to measure the generated down-converted photons \cite{Optica_2020}. The design is flexible, since it allows for the core geometry of the fibers to be changed without realignment of the source, a crucial feature since MCF technology is in a rapid stage of development, and widespread standards have yet to be established. To demonstrate its viability, we prepare an entangled state of two four-dimensional systems, encoded in the path degree of freedom of the down-converted photons, and certify the multi-dimensional entanglement generation. In addition, quantum steering is identified, indicating entanglement in a one-sided device-independent scenario.

The source presents several technical advantages, including high spectral brightness - comparable with modern polarization-entangled-qubit sources, and relatively long phase stability, thanks to the use of multi-core fibers. Consequently, our scheme has several potential applications. For instance, one can exploit the verified multi-dimensional steering for implementing one-side device independent quantum protocols such as quantum cryptography or quantum randomness generation \cite{branciard12,law14,passaro15,mattar17,Gomez_2018}. The spectral brightness achieved of 350000 photon pairs generated (s mW nm)$^{-1}$, will allow for future investigations demonstrating the viability of long-distance distribution of multi-dimensional entanglement over multi-core fibers. Last, we note that the presented scheme is readily available for use in quantum-enhanced metrology beyond the standard quantum limit. In particular, the entangled state produced by our source is nearly optimal for multi-parameter phase estimation \cite{Fabio_2019}.

\section*{Acknowledgements}
We thank T. García and J. Cariñe for laboratory assistance, and M. Huber and G. Xavier for providing valuable comments. This work was supported by Fondo Nacional de Desarrollo Cient\'{i}fico y Tecnol\'{o}gico (FONDECYT) (1190901, 1200266, 1200859) and Millennium Institute for Research in Optics. S.G. acknowledges ANID. AC was supported by Universidad de Sevilla Project Qdisc (Project No.\ US-15097), with FEDER funds, MINECO Project No.\ FIS2017-89609-P, with FEDER funds, and QuantERA grant SECRET, by MINECO (Project No.\ PCI2019-111885-2). S.P. thanks CNPq - Conselho Nacional de Desenvolvimento Cient\'ifico
e Tecnol\'ogico and INCT-IQ - Instituto Nacional de Ci\^encia e Tecnologia de Informac\~ao Qu\^antica. S. P. express gratitude to the University of Concepci\'on.

\section*{Appendix: Fidelity Calculation from Probability Distributions}
The fidelity was calculated from the experimental data by defining the Pauli operators for the $j,k$ subspace as \cite{Padua_2014,bavaresco18}: $\sigma^{(jk)}_x=\ket{j}\bra{k}+\ket{k}\bra{j}$, $\sigma^{(jk)}_y=i\ket{j}\bra{k}-i\ket{k}\bra{j}$. Then,  $\mathrm{Re} [ \langle jj | \rho | kk \rangle ] = (\langle \sigma^{(jk)}_x \otimes \sigma^{(jk)}_x\rangle - \langle \sigma^{(jk)}_y \otimes \sigma^{(jk)}_y \rangle)/4$.
Let us denote the probabilities from the $X_j$ measurements as $P^{(j)}=\{ P^{(j)}_{xy} \}$, with $x,y=0,\dots3$.  We also define the correlation functions
\begin{align}
C_j(\alpha,\beta,\gamma,\delta) = & [ P^{(j)}_{\alpha\alpha}+ P^{(j)}_{\alpha\beta}+ P^{(j)}_{\beta\alpha}+P^{(j)}_{\beta\beta}  \\ \nonumber
+ & P^{(j)}_{\gamma\gamma}+P^{(j)}_{\gamma\delta}+P^{(j)}_{\delta\gamma}+P^{(j)}_{\delta\delta} \\ \nonumber
- & P^{(j)}_{\alpha\gamma}-P^{(j)}_{\alpha\delta}-P^{(j)}_{\alpha\gamma}-P^{(j)}_{\alpha\delta} \\ \nonumber
 - & P^{(j)}_{\gamma\beta}-P^{(j)}_{\gamma\beta}-P^{(j)}_{\delta\beta}-P^{(j)}_{\delta\beta} ], \\
\end{align}
for $\alpha,\beta,\gamma,\delta=0,1,2,3$. Then, it is a straightforward but lengthly calculation to show that we can use the $X_j$ measurement probability distributions in the above correlation function to determine the terms in the Fidelity.  Explicitly, we have
 \begin{align}
2  \mathrm{Re} [ \langle 00 | \rho | 11 \rangle ] + 2  \mathrm{Re} [ \langle 22 | \rho | 33 \rangle ] =  \\ \nonumber
\frac{1}{4} [C_{0}(0,1,2,3)+C_{1}(0,1,2,3) \\ \nonumber
 -C_{2}(0,1,2,3)-C_{3}(0,1,2,3)],
 \end{align}
  \begin{align}
2  \mathrm{Re} [ \langle 00 | \rho | 22 \rangle ] + 2  \mathrm{Re} [ \langle 11 | \rho | 33 \rangle ]  = \\ \nonumber  \frac{1}{4} [ C_{0}(0,2,1,3)+C_{1}(0,2,1,3) \\ \nonumber
 -C_{2}(0,2,1,3)-C_{3}(0,2,1,3)],
 \end{align}
   \begin{align}
2  \mathrm{Re} [ \langle 00 | \rho | 33 \rangle ] + 2  \mathrm{Re} [ \langle 11 | \rho | 22 \rangle ] = \\ \nonumber  \frac{1}{4} [ C_{0}(0,3,1,2)+C_{3}(0,3,1,2) \\ \nonumber
 -C_{1}(0,3,1,2)-C_{2}(0,3,1,2)].
 \end{align}
 These expressions can be used directly in Eq. \eqref{eq:Fid} to calculate the fidelity.

%%% BIBLIOGRAPHY %%%%%%%%%%

\end{document}